\documentclass[conference]{IEEEtran}
\usepackage[dvipsnames]{xcolor}
\usepackage{cite}
\usepackage{amsmath,amssymb,amsfonts}
\usepackage{algorithmic}
\usepackage{graphicx}
\usepackage{textcomp}
\usepackage{lipsum}
\usepackage{adjustbox}
\usepackage{booktabs}
\usepackage{pifont}
\usepackage{xspace}
\usepackage{tikz}
\def\BibTeX{{\rm B\kern-.05em{\sc i\kern-.025em b}\kern-.08em
    T\kern-.1667em\lower.7ex\hbox{E}\kern-.125emX}}

\newcommand{\sym}[1]{{#1}}

\newcommand{\name}{\sym{PacQ}\xspace}

\newcommand*\colourcheck[1]{%
  \expandafter\newcommand\csname #1check\endcsname{\textcolor{#1}{\ding{52}}}%
}

\newcommand*\colourcross[1]{%
  \expandafter\newcommand\csname #1cross\endcsname{\textcolor{#1}{\ding{56}}}%
}
\colourcheck{teal}
\colourcross{red}

\newcommand*\circled[1]{\tikz[baseline=(char.base)]{
            \node[shape=circle,fill,inner sep=0.8pt] (char) {\textcolor{white}{#1}};}}
\begin{document}

\title{PacQ: A SIMT Microarchitecture for Efficient Dataflow in Hyper-asymmetric GEMMs}

\author{\IEEEauthorblockN{Ruokai Yin}
\IEEEauthorblockA{
\textit{Yale University}\\
ruokai.yin@yale.edu}
\and
\IEEEauthorblockN{Yuhang Li}
\IEEEauthorblockA{
\textit{Yale University}\\
yuhang.li@yale.edu}
\and
\IEEEauthorblockN{Priyadarshini Panda}
\IEEEauthorblockA{
\textit{Yale University}\\
priya.panda@yale.edu}
}

\maketitle

\begin{abstract}
Weight-only quantization has been widely explored in large language models (LLMs) to reduce memory storage and data loading overhead. During deployment on single-instruction-multiple-threads (SIMT) architectures, weights are stored in low-precision integer (INT) format, while activations remain in full-precision floating-point (FP) format to preserve inference accuracy. Although memory footprint and data loading requirements for weight matrices are reduced, computation performance gains remain limited due to the need to convert weights back to FP format through unpacking and dequantization before GEMM operations. In this work, we investigate methods to accelerate GEMM operations involving packed low-precision INT weights and high-precision FP activations, defining this as the hyper-asymmetric GEMM problem. Our approach co-optimizes tile-level packing and dataflow strategies for INT weight matrices. We further design a specialized FP-INT multiplier unit tailored to our packing and dataflow strategies, enabling parallel processing of multiple INT weights. Finally, we integrate the packing, dataflow, and multiplier unit into \name, a SIMT microarchitecture designed to efficiently accelerate hyper-asymmetric GEMMs. We show that \name can achieve up to $1.99\times$ speedup and $81.4\%$ reduction in EDP compared to weight-only quantized LLM workloads running on conventional SIMT baselines.
\end{abstract}

\section{Introduction}
% Large language models (LLMs) have achieved significant performance improvements across a wide range of complex natural language processing tasks~\cite{zhang2022opt, touvron2023llama}. However, the huge amount of parameters accompanied by the LLMs has brought deployment challenges such as increased memory footprint requirements and computational costs. One of the popular solutions for enhancing the deployment efficiency of LLMs on the edge is quantization. Conventional DNN quantization involves quantizing both the weights and activations to low-precision integers to reduce both the computation and memory traffic of the models~\cite{li2021brecq}. However, quantizing activations in LLMs poses significant challenges due to their dynamic range and the small portion of salience~\cite{lin2024awq,xiao2023smoothquant}. Consequentially, many prior works focus on the weight-only quantization for LLMs, where the weights are compressed towards very low-precision integer ($\leq$INT4), while the activations are kept as high precision floating point format (usually FP16)~\cite{frantar2022gptq,lin2024awq,xiao2023smoothquant,wang2023bitnet,li2024tesseraq}.

Large language models (LLMs) have demonstrated exceptional performance across a wide range of complex natural language processing tasks~\cite{zhang2022opt, touvron2023llama}. However, their substantial parameter counts introduce significant deployment challenges, including increased memory footprint and computational costs. Quantization has emerged as a popular solution to improve the deployment efficiency of LLMs, particularly on edge devices.

Conventional deep neural network (DNN) quantization reduces both computation and memory traffic by compressing weights and activations into low-precision integers~\cite{li2021brecq,jacob2018quantization}. However, quantizing activations in LLMs is particularly challenging due to their dynamic range and the occurrence of salient values~\cite{lin2024awq,xiao2023smoothquant}. As a result, many recent studies have focused on weight-only quantization, where weights are compressed to very low-precision integers (e.g., $\leq$INT4), while activations are preserved in high-precision floating-point formats, typically FP16~\cite{frantar2022gptq,lin2024awq,xiao2023smoothquant,wang2023bitnet,li2024tesseraq}.

%%%% Figure 1 %%%%
\begin{figure}[t]
\centering
\includegraphics[width=\linewidth]{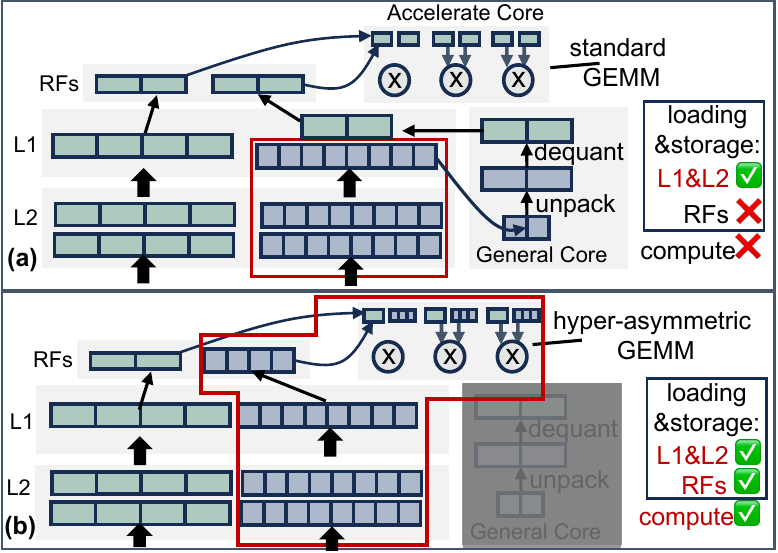}
  % \caption{(a) Standard inference flow on SIMT architecture for weight-only quantized LLMs. \textcolor[HTML]{AED9BF}{\textbf{Green}} denotes high-precsion floating point activations, and \textcolor[HTML]{ADB9EA}{\textbf{blue}} denotes low-precision integer weights. (b) Our proposed flow leverages the hyper-asymmetric GEMM to enjoy the loading \& storage as well as compute benefits of low-precision integer weights throughout the entire stack of GEMM computations.}
  \caption{(a) Standard inference flow on SIMT architecture for weight-only quantized LLMs. \textcolor[HTML]{AED9BF}{\textbf{Green}} denotes high-precision floating-point activations, and \textcolor[HTML]{ADB9EA}{\textbf{blue}} denotes low-precision integer weights. (b) Our proposed flow leverages the hyper-asymmetric GEMM to achieve the benefits of reduced loading, storage, and compute costs from low-precision integer weights throughout the entire GEMM computation stack.}
  \label{fig:intro}
  \vspace{-5mm}
\end{figure}
%%%% Figure 1 %%%%
% Though quantizing the weights substantially reduces the memory traffic of the LLMs (e.g., Llama2-70B~\cite{touvron2023llama} with FP16 weights requires 131.6 GB memory but only 35.8 GB with INT4 weight-only quantization), the computation cost of those models is not saved. The reason lies in the current deployment flow of those models on single-instruction multiple-thread (SIMT) based hardware, which is widely adopted by commercial ML accelerators (e.g., NVIDIA's GPU). As illustrated in Figure.~\ref{fig:intro}(a), the current inference deployment flow of weight-only quantized LLMs on SIMT hardware has three major limitations: \textbf{First}, though the weights in low precision integer (INT) will be stored and fetched in packed format thus leading to reduced memory storage and traffic, this reduction only happens in the memory hierarchy below the L1 cache. Once the packed data are loaded into the L1 cache, they will be sent to the general core for unpacking and dequantized to the high precision floating point (FP) and stored back into the L1 cache. \textbf{Second}, the process of unpacking and dequantization poses non-negligible overheads~\cite{lin2024awq}. \textbf{Third}, after the dequantization, the computation eventually happens in FP, thus leading to no reduction in the computation compared to the FP models.
While weight quantization significantly reduces memory traffic for LLMs (e.g., Llama2-70B~\cite{touvron2023llama} requires 131.6 GB with FP16 weights but only 35.8 GB with INT4 weight-only quantization), it does not lower computational costs. This limitation stems from the current deployment workflow on single-instruction multiple-thread (SIMT) hardware, widely used in commercial ML accelerators such as NVIDIA GPUs. As shown in Figure~\ref{fig:intro}(a), the inference deployment flow of weight-only quantized LLMs on SIMT hardware faces three major challenges: (1) \textbf{\textit{Inefficient memory hierarchy utilization:}} Although low-precision integer (INT) weights are stored and fetched from off-chip DRAM in packed format, they are unpacked and dequantized into high-precision floating-point (FP) format upon loading into the L1 cache, losing memory benefits at and above the L1 cache.
(2) \textbf{\textit{Overhead of unpacking and dequantization:}} Unpacking and dequantizing weights introduce significant latency and computational overhead~\cite{lin2024awq}.
(3) \textbf{\textit{Lack of computational savings:}} GEMM computations are performed in FP after dequantization, forfeiting the computational efficiency of low-precision integer weights.

% Despite the limitations, existing weight-only quantization work usually achieve inference speedup on SIMT architecture in memory-bounded scenarios, for example, the single-batch text generation. However, multi-batch processing is widely adopted in real-world LLM serving systems for its better utilization of hardware~\cite{yu2022orca}. The multi-batch processing of LLMs is in general a compute-bounded workload, whose performance is limited again by the three limitations. 
Despite these limitations, existing weight-only quantization techniques often achieve inference speedup on SIMT architectures in memory-bound scenarios, such as single-batch text generation. However, real-world LLM serving systems predominantly adopt multi-batch processing to achieve better hardware utilization~\cite{yu2022orca}. Multi-batch processing, in contrast, is typically compute-bound, with its performance constrained by the three limitations outlined earlier.

% To overcome the limitations in the deployment of weight-only quantized models on SIMT architecture, we propose to keep the INT weights in packed form until the computation completes. We define this problem as \textbf{hyper-asymmetric GEMM} where the precision of two operands are extremely unbalanced (asymmetric) for the GEMM operation, as shown in Figure.~\ref{fig:intro}(b). In this work, we explore the design space of accelerating the hyper-asymmetric GEMM on SIMT-based hardware. Our key contributions are:
To address the limitations in deploying weight-only quantized models on SIMT architectures, we propose to retain INT weights in packed format throughout the entire GEMM computation stack. We define this problem as \textbf{hyper-asymmetric GEMM}, where the operand precisions are highly unbalanced (asymmetric), as illustrated in Figure~\ref{fig:intro}(b). In this work, we explore the design space for accelerating hyper-asymmetric GEMM on SIMT-based hardware. Our key contributions are:

% (1) We pinpoint the importance of the packing direction of INT weights, which is often neglected in prior works, for hyper-asymmetric GEMM. We analysis the inefficiency of packing the weights along the input-feature ($k$) dimension which is widely adopted in existing quantized LLM deployment frameworks~\cite{autogptq,llmc}. We propose an efficient packing and dataflow for hyper-asymmetric GEMM. Results indicate that our proposed packing and dataflow reduces the number of register file accesses by up to $x \times$ compared to baselines.
 \noindent (1) We pinpoint the critical role of INT weight packing direction, a factor often overlooked in prior work, for optimizing hyper-asymmetric GEMM. Specifically, we analyze the inefficiencies of packing weights along the input-feature ($k$) dimension, a common approach in existing quantized LLM deployment frameworks~\cite{autogptq,llmc}. To address this, we propose an alternative packing strategy and dataflow optimized for hyper-asymmetric GEMM. Our strategy has 54.3$\%$ reduction of register file accesses compared to baseline approaches.

% (2) We propose a parallel FP-INT multiplier that works closely with our proposed packing and dataflow. We observe constant patterns in the FP representation of the INT weight values. By leveragin the pattern, our design can process the 4 FP16$\times$INT4 and 8 FP16$\times$INT2 multiplications in parallel in one cycle by reusing most of the hardware resources in the standard FP16 multiplier design. Our design only incurs $x\%$ extra area and power to the conventional design while improve the throughput by $x\times$.
\noindent (2) We propose a parallel FP-INT multiplier optimized for our packing and dataflow strategy. By identifying constant patterns in the FP representation of INT weight values, our design efficiently processes four FP16$\times$INT4 or eight FP16$\times$INT2 multiplications in parallel within a single cycle, reusing $\sim$73\% of hardware resources from standard FP16 multipliers. This design achieves up to 6.8$\times$ throughput/watt improvement compared to conventional designs.

% (3) We encapsulate our packing and dataflow togethr with the parallel multiplier design in \name, a SIMT microarchitecture for hyper-asymmetric GEMM. Compared to our baseline of NVIDIA's GPUs, \name achieve $x \times$ throughput improvements and $y\times$ inference energy reduction on selected LLM workloads.
\noindent (3) We encapsulate the proposed packing, dataflow, and parallel multiplier design into \name, a SIMT microarchitecture tailored for hyper-asymmetric GEMM. Compared to NVIDIA GPUs as a baseline, \name achieves up to $81.4\%$ EDP reduction for selected LLM inference workloads.

\section{Background}

\noindent \textbf{SIMT Architecture.}\indent The single-instruction multiple-thread (SIMT) architectures are widely used in commercial ML accelerators, such as NVIDIA GPUs, to achieve high computational, memory, and control parallelism via sets of consecutively indexed threads (warps). These architectures tile large ML workloads (e.g., GEMMs) into smaller blocks mapped to individual threads, improving bandwidth efficiency and data reuse~\cite{tan2011fast}. NVIDIA’s Volta GPUs introduced the first streaming multiprocessors (SMs) with dedicated ML acceleration capabilities, including tensor cores (TCs), specialized for GEMM acceleration~\cite{v100whitepaper, demystifyvolta}.

In Volta, multiple SMs are interconnected via an on-chip network to a last-level cache backed up by HBM. Each SM contains sub-cores sharing an L1 cache, with sub-cores further divided into general-purpose cores and tensor cores. General-purpose cores handle operations like unpacking and dequantization, while tensor cores are optimized for GEMM operations. Operand data is stored in register files for efficient reuse before being processed by tensor cores. Without loss of generality, our study examines the dataflow and microarchitecture of Volta’s tensor cores as the baseline.

\begin{figure}[t]
\centering
\includegraphics[width=0.9\linewidth]{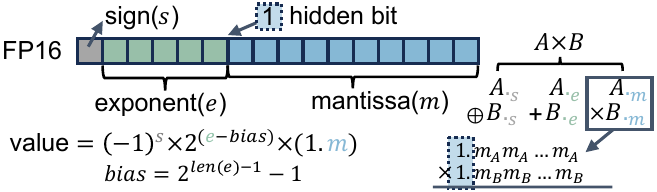}
\vspace{-3mm}
  \caption{IEEE 754 standard FP16 format.}
  \label{fig:bg:fp}
  \vspace{-5mm}
\end{figure}

% \subsection{Floating Point Number}
\noindent \textbf{Floating Point Number.}\indent In Figure~\ref{fig:bg:fp}, we show the standard IEEE 754 format for FP16. In this work, we focus on the normalized format (i.e., the hidden bit for the mantissa is always 1). During the floating point multiplication, the hidden bit needs to be considered for the integer multiplication between the mantissas, as shown on the right side of Figure~\ref{fig:bg:fp}.
%%%% Figure 1 %%%%
%%%% Figure 1 %%%%
\noindent \textbf{Related Work.}\indent Previous studies~\cite{reggiani2023mix,gope2020high} on accelerating mixed-precision GEMM for quantized DNNs have primarily targeted workloads with low operand precision imbalance (difference $\leq4\times$). In contrast, our work focuses on hyper-asymmetric GEMM, characterized by a high operand precision imbalance (difference$ \geq4\times$).

Recent work, FIGNA~\cite{jang2024figna}, introduced an FP-INT unit to eliminate unpacking and dequantization overhead during the deployment of weight-only quantized LLMs. However, their approach emphasizes co-designing the FP-INT multiplier with a block floating-point format. In contrast, our work integrates the FP-INT multiplier design with packing and dataflow strategies for processing packed INT weights. Furthermore, our design improves throughput by enabling parallel multiplication of FP activations with multiple INT weights, a capability not explored in FIGNA.

Other methods for accelerating mixed-precision GEMM in weight-only quantized LLMs employ lookup table (LUT) techniques, either by designing new LUT-friendly hardware~\cite{mo2024lut} or focusing on single-batch text generation on GPUs~\cite{park2022lut}. In contrast, our work enhances general SIMT-based architectures to support multi-batch inference, a scenario more common in real-world LLM deployments. Additionally, approaches targeting efficient dequantization kernels~\cite{lin2024awq,zhao2024atom} address complementary challenges and are orthogonal to our work.

%%%% Figure 2 %%%%
\begin{figure*}[t]
\centering
\includegraphics[width=0.85\linewidth]{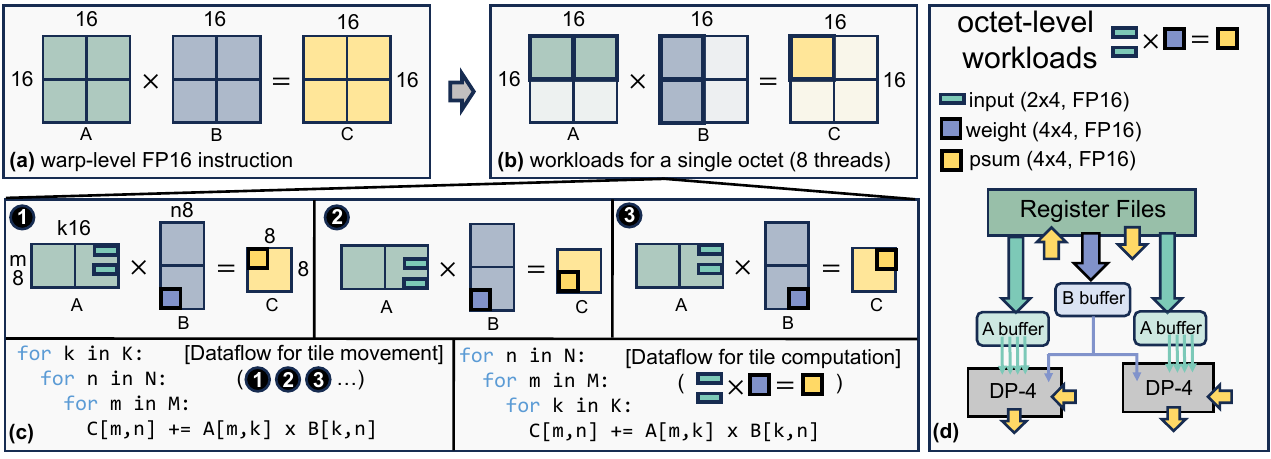}
  % \caption{(a) \& (b): The tile mapping of the warp-level (a set of 32 threads) \textit{mma.sync.m16n16k16} instruction to different octets (a subset of 8 threads). Here, $A$ denotes the input matrix with the shape of [$m,k$], $B$ denotes the weight matrix with the shape of [$k,n$], and $C$ denotes the output (partial-sum) matrix with the shape of [$m,n$]. (c) Example of a step-by-step breakdown of the octet-level data movement and computation flow. (d) Mapping of octet-level workloads into the baseline's tensor processing units.}
  \vspace{-3mm}
  \caption{(a) \& (b): The tile mapping of the warp-level (a set of 32 threads) \textit{mma.sync.m16n16k16} instruction to different octets (a subset of 8 threads). Here, $A$ represents the input matrix with a shape of [$m,k$], $B$ represents the weight matrix with a shape of [$k,n$], and $C$ represents the output (partial-sum) matrix with a shape of [$m,n$]. (c) An example of a step-by-step breakdown of the octet-level data movement and computation flow. (d) The mapping of octet-level workloads onto the baseline tensor processing units. DP-4 stands for four-element dot-product units that performs $4\times4$ inner-product.}
  \vspace{-3mm}
  \label{fig:dataflow:baseline}
\end{figure*}

\section{Packing flow for Hyper-asymmetric GEMM}
% In this section, for a better understanding of the existing SIMT architecture's limitation on precessing asymmetric-GEMM, we set up the NVIDIA-Volta~\cite{v100whitepaper} GPU microarchitecture as our baseline. We first describe Volta's dataflow for the asymmetric GEMM (when the weight matrices are packed). And then explain the associated problem. Finally, we describe our solution.
% We describe the Volta GPU's dataflow of running the FP16 tensor core MMA instruction of \textit{mma.sync.m16n16k16} for \textit{W16A16} to demonstrate the standard GEMM dataflow on SIMT architecture, as shown in Figure.~\ref{fig:dataflow:baseline}. In the provided example, each matrix is $16\times16$, and \textit{W16A16} means the network has both weights and activations in FP16 precision.
%
\noindent \textbf{Basic GEMM Dataflow.}\indent For illustration, we describe the Volta GPU's execution of the FP16 tensor core MMA instruction \textit{mma.sync.m16n16k16} for \textit{W16A16}, where both weights and activations are in FP16 precision, as shown in Figure~\ref{fig:dataflow:baseline}. In this example, each matrix has a size of 16$\times$16.
As shown in Figure~\ref{fig:dataflow:baseline}(a), matrices $A$, $B$, and $C$ are collaboratively fetched by a warp consisting of 32 threads. 
%
% Then, as shown in Figure.~\ref{fig:dataflow:baseline}(b), the entire workload will be evenly distributed to 4 different \textit{octets} (8 threads).
The workload is then evenly distributed among four \textit{octets}~\cite{demystifyvolta} (groups of 8 threads), as illustrated in Figure~\ref{fig:dataflow:baseline}(b).
%
% Within each \textit{octet}, the workload is further broken down into smaller tiles as shown in Figure.~\ref{fig:dataflow:baseline}(c). Each octet will fetch the tiles following the weight-stationary dataflow described by the left 3-nested-for-loop. The computation is done within each tile through the output-stationary dataflow described by the 3-nested-for-loop on the right.
As depicted by Figure~\ref{fig:dataflow:baseline}(c), each \textit{octet} fetches tiles using a weight-stationary dataflow, represented by the left three-nested-for-loop, and computations within each tile follow an output-stationary dataflow, represented by the right three-nested-for-loop.
Finally, Figure~\ref{fig:dataflow:baseline}(d) illustrates how octet-level workloads are mapped onto the hardware (i.e., tensor cores in Volta GPUs). A $4\times4$ tile of $B$ is fetched from the register files into a buffer shared by all eight threads within the \textit{octet}. Meanwhile, $2\times4$ tiles of $A$ are loaded into separate buffers, each shared by four threads. Two four-element dot-product units (DP-4) are responsible to compute the $4\times4$ tile of $C$.

\noindent \textbf{Hyper-asymmetric GEMM Packing and Dataflow.}\indent We define hyper-asymmetric GEMM as a GEMM operation where the operands $A$ and $B$ have a precision difference of at least $4\times$ (e.g., \textit{W4A16} or \textit{W2A16}). While largely overlooked in prior dequantization-based GEMM methods, the dimension along which the INT weights are packed becomes a critical factor in hyper-asymmetric GEMM. Formally, we define the packing of weight matrix $B$ using the format $P(B_x)_y$, where $x$ represents the number of elements of $B$ packed along the $y$ dimension. For instance, in \textit{W4A16}, if the weights ($B$) are packed along the $k$ dimension into the INT16 datatype, the packing can be formally described as $P(B_4)_k$. Once fetched to the L1 cache, the INT16 is unpacked to 4 INT4 weights and dequantized to FP16 for being executed in the standard FP16 GEMM. 

Although there is limited research on the packing direction of INT weight matrices, nearly all prior weight-only LLM quantization frameworks~\cite{llmc,autogptq,frantar2022optq} choose to pack weights along the input feature dimension ($k$-dim). While the packing dimension is unimportant in dequantization-based GEMM, as packed INT weights are unpacked at the L1 cache regardless of the packing direction, this is not the case for hyper-asymmetric GEMM. In hyper-asymmetric GEMM, INT operands are fetched in their packed form, creating hyper-asymmetry in the amount of data staged in the registers and tensor cores. An improper packing dimension combined with this hyper-asymmetry leads to significant performance degradation. 

\begin{figure}[t]
\centering
\includegraphics[width=0.9\linewidth]{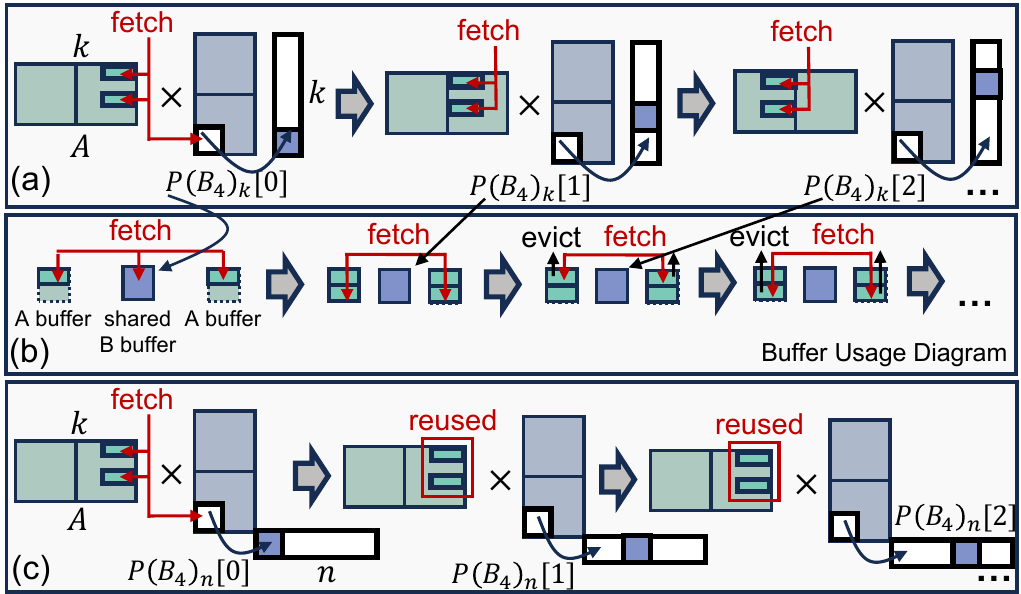}
% \vspace{-3mm}
  % \caption{(a) Illustration of multiple fetch instructions required by $k$ dimension packing. (b) Illustration of the poor data reuse of $A$ in $k$ dimension packing. (c) Illustration of the benefits of packing along $n$ dimension. The tile size is identical to Figure.~\ref{fig:dataflow:baseline}.}
\caption{(a) Illustration of multiple fetch instructions required by the hyper-asymmetric GEMM when packing weights along the $k$ dimension. (b) Illustration of the poor data reuse of $A$ in $k$ dimension packing. (c) Illustration of the improved data reuse and reduction in fetch instructions achieved by packing along the $n$ dimension. The tile size is identical to Figure.~\ref{fig:dataflow:baseline}.}
  \vspace{-5mm}
  \label{fig:dataflow:packin} 
\end{figure}

Specifically, when weights are packed along the $k$-dim, each packed INT weight fetch requires multiple activation fetch instructions because the $k$-dim must align between the two operands during GEMM computation. Figure~\ref{fig:dataflow:packin}(a) provides an example: for processing INT weights packed in the $P(B_4)_k$ format, four distinct fetch instructions for $A$ must be issued to align the operands.
Furthermore, packing INT weights ($B$) along the $k$-dim results in poor data reuse of activations ($A$). As illustrated in Figure~\ref{fig:dataflow:packin}(b), new data from $A$ along the $k$-dim must be continuously fetched, leading to eviction of buffers inside the tensor cores and preventing data reuse of $A$. This causes excessive data access from large register files, which is highly power-intensive. When the operand precision asymmetry is significant, this issue can even escalate beyond the register file level to the L1 cache.

%%%% Figure 1 %%%%
%%%% Figure 1 %%%%
\begin{figure*}[t]
\centering
\includegraphics[width=0.8\linewidth]{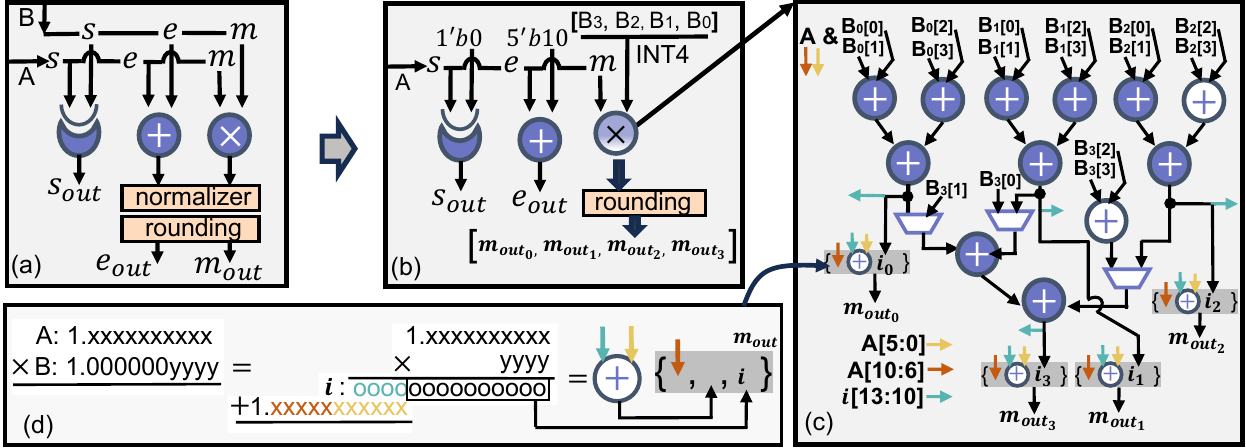}
\vspace{-3mm}
  \caption{(a) Standard FP multiplier design. $s$ stands for sign, $e$ represents exponent, and $m$ denotes the mantissa. (b) Overview of our proposed parallel FP-INT-16 multiplier. In the provided example, our design can generate four output mantissas in one cycle. (c) Detailed diagram of the modifications made to the original 11-bit integer multiplier. The elements in \textcolor[HTML]{6C74C4}{purple} are part of the original multiplier, while the elements in white represent additional units. All bits of $B$'s mantissas perform a logical AND operation with shifted $A$'s mantissas before being fed into the adders. (d) Explanation of how the final output mantissa (shaded gray) is assembled from different value sources. $\{\}$ denotes the append operation.}
  \label{fig:mul_design}
\end{figure*}

We argue that INT weights should instead be packed along the output feature ($n$) dimension. This approach avoids additional internal loops of fetch instructions compared to processing INT weights packed along the $k$-dim. As shown in Figure~\ref{fig:dataflow:packin}(c), once the activations ($A$) and packed weights ($P(B_4)_n$) are fetched following the standard flow, no extra fetch instructions are required. Furthermore, our proposed packing strategy ensures no eviction of $A$ during the processing of packed $B$, enabling full reuse of activations across the packed weights. Moreover, instead of using the weight stationary dataflow for tile movement as in Figure~\ref{fig:dataflow:baseline}(c), we use output stationary for both tile movement and tile computation. This packing and dataflow combination strikes a good balance between the data reuse and the partial sum traffic.

% Since the packing of the weight tensors is usually done in the front-end of the deep learning framework (e.g., PyTorch), there is 

% As illustrated in Figure.~\ref{fig:dataflow:baseline}(c), the computation of a tile follows the output-stationary dataflow, where the dot product is performed along the $k$ dimension. If the computation stage of hyper-asymmetric GEMM is using the standard output stationary dataflow, the INT weights have to be packed along the $k$ dimension of the weight matrix. The reason is that

\section{Proposed Parallel FP-INT Multiplier}
\label{sec:fpint-mul}
%%%% Figure 4 %%%%
%%%% Figure 4 %%%%

As discussed previously, our proposed packing and data flow reuses activations across all packed INT weights within a single data fetch instruction. In the standard approach, these data are dequantized and processed sequentially by the pipelined FP multiplier. However, we identify an opportunity to perform the multiplication between a single FP activation and multiple INT weights in parallel. To exploit this, we propose a specialized parallel FP-INT multiplier.

\noindent \textbf{Observation from FP representation of INT values.}\indent When representing an INT $x\in[1024, 2048)$ in FP16 format, certain patterns emerge during the conversion process. First, the FP16 representation of $x$ consistently has an exponent value of $11001$ (corresponding to $2^{25-15} = 1024$ in decimal). Second, the mantissa of $x$ is always in the form of $10'b0 | (x - 1024)$. Based on these patterns, we make further observations on representing low-precision INT weights in FP16 format. Without loss of generality, we use INT4 as an example, which can be easily extended to INT2~\cite{kim2022says}.

Assume $B \in [-8, 7)$ is a signed INT4 weight scalar. To simplify arithmetic operations, we first transform $B$ to an unsigned representation by adding 8, resulting in $B+8 \in [0, 15)$. Since $B$ is in INT4 format, we ensure that $B+8+1024$ ($B+1032$) lies within the range $[1024, 2048)$. Leveraging the observed patterns, we conclude: \circled{1} The FP16 representation of $B+1032$ always has an exponent value of $11001$. \circled{2} The mantissa of $B+1032$ is consistently in the form $000000yyyy$, where $yyyy$ is the 4-bit integer representation of $B+8$.

% \subsection{Proposed Design}

\noindent \textbf{Proposed Design.}\indent Leveraging \circled{1} and \circled{2}, we design a parallel FP-INT multiplier that maximally reuses the hardware resources of the original FP multiplier, as shown in Figure~\ref{fig:mul_design}(a). Our design enables the multiplication of one FP16 input with four INT4 values or eight INT2 values in parallel, producing all outputs in a single cycle. For illustration, we focus on INT4.

Since $B$ is transformed into an unsigned representation, the signs of all four outputs depend only on the sign of $A$. Thus, we compute the common output sign $s_{out}$ by XOR-ing $A$'s sign bit with $0$. Using \circled{1}, we derive the shared exponent $e_{out}$ for all outputs by adding $A$'s 5-bit exponent to $11001$, as shown in Figure~\ref{fig:mul_design}(b).
Figure~\ref{fig:mul_design}(c) depicts the details of our proposed parallel FP-INT multiplier. The original FP16 integer multiplier requires 10 parallel adders (16-bit) to compute one 11-bit$\times$11-bit multiplications (colored \textcolor[HTML]{6C74C4}{purple}). Utilizing \circled{2}, we break the original four 11-bit$\times$11-bit multiplications into four 11-bit$\times$4-bit multiplication. By introducing two additional 16-bit parallel adders (colored white), our design simultaneously performs all four 11-bit$\times$4-bit multiplications.

Finally, as shown in Figure~\ref{fig:mul_design}(d), we ensure the hidden bit of $B$'s mantissa is accounted for by performing a 6-bit addition between the top-4 MSBs of the intermediate results $i$ (colored \textcolor[HTML]{58b4b0}{turquoise}, from 11-bit$\times$4-bit) and the 6 LSBs of $A$ (colored \textcolor[HTML]{e8c65f}{yellow}). We then concatenate the top-5 MSBs of $A$ (colored \textcolor[HTML]{c55a11}{brown}) with the addition results and $i$ to obtain the four final mantissa products $m_{out}$. These results are passed to the rounding units and truncated to 10 bits (excluding the hidden bit). Since both $A$ and $B$ have normalized mantissa and $B$'s mantissa values are constrained to a 4-bit unsigned integer, normalization to the outputs is unnecessary.

%%%% Figure 1 %%%%
\begin{figure}[t]
\centering
\includegraphics[width=0.8\linewidth]{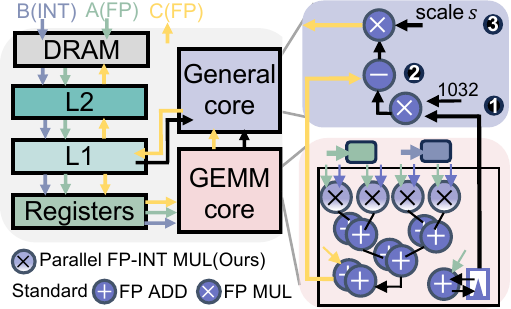}
  \caption{Overview of the \name architecture. $B$ represents the INT weight vectors, $A$ represents the FP activation vectors, and $C$ represents the outputs or partial sums in FP16. The steps illustrated in the general core reflect only the required instructions in sequence. \name does not require any hardware modifications to the general core.}
  \label{fig:arch}
  \vspace{-5mm}
\end{figure}
%%%% Figure 1 %%%%

\noindent \textbf{Overall Architecture (PacQ).}\indent We encapsulate our proposed packing and dataflow strategies along with the parallel FP-INT multiplier units into \name: a SIMT microarchitecture for accelerating hyper-asymmetric GEMMs. The overall architecture is depicted in Figure~\ref{fig:arch}. Our proposed microarchitecture retains most components from the Volta GPU baseline, requiring hardware modifications only in the GEMM acceleration core (i.e., tensor core). We introduce the following three key changes to the original tensor core:

\textbf{First}, we replace all original FP16 multipliers with our proposed parallel FP-INT multipliers. \textbf{Second}, we duplicate the original adder trees in the DP-4 by twice, enabling the accumulation of the inner product of 16 values in 2 cycles for INT4 weights (or 32 values in 4 cycles for INT2 weights). \textbf{Third}, we equip the GEMM core with small accumulators to store processed $A$ values. Recall that before feeding the packed INT weights into the parallel FP-INT multipliers, we transform all weight values $B$ into $B+1032$. To retrieve the original weight values, we subtract $1032$ later in the computation. By leveraging the small accumulators, we fuse this subtraction directly into the inner product calculation:
\begin{equation}
    \sum_0^k(A_k(B_k-1032))=\sum_0^kA_kB_k - 1032\times\sum_0^kA_k,
\end{equation}
where $k$ denotes the dimension of the inner product. The accumulators are responsible for generating the results of $\sum_0^k A_k$. As shown in the general core of Figure~\ref{fig:arch}, the accumulated results are multiplied by $1032$ in \circled{1}, then subtracted from the inner product results ($\sum_0^k A_k B_k$) in \circled{2}. Finally, the group quantization scales $s$ are applied in \circled{3} to scale the outputs back to their expected range.

\section{Experimental Results}

% \subsection{}

% \noindent \textbf{Experimental Setup.}\indent For the packing and dataflow evaluation, we have written our own simulator in Python to monitor the memory access pattern and to record the running cycles for the GPU baseline architectures. We implemented the key components of \name and our baselines in RTL and synthesize them using Synopsys Design Compiler at 400MHz with 32nm technology. We use CACTI7.0~\cite{cacti} to model the on-chip SRAM to obtain memory statistics. The configuration details of \name and our baselines can be found in Table~\ref{tb:hw_config}. If not specified, the INT2 and INT4 refer to the precision of weights. And the activation and partial-sum is always FP16.
\noindent \textbf{Experimental Setup.} \indent For the packing and dataflow evaluation, we developed a custom simulator in Python to monitor memory access patterns and record execution cycles for \name and the baselines. Key components of \name and our baselines were implemented in RTL and synthesized using Synopsys Design Compiler at 400MHz with 32nm technology. We utilized CACTI7.0~\cite{cacti} to model on-chip SRAM and register files for getting memory statistics. Configuration details for \name and the baseline architectures are summarized in Table~\ref{tb:hw_config}. Unless otherwise specified, in the rest of the section, INT2 and INT4 refer to the precision of weights, while activations and partial sums are always represented in FP16 format. For hardware unit comparisons, 'baseline' always refers to the standard baseline designs as denoted in Table.~\ref{tb:hw_config}. $P(B_{4(8)})_k$ denotes the baseline of hyper-asymmetric GEMM flow that packs 4 INT4 (8 INT2) weights in one INT16 along $k$-dim. 
%We are not showing the statistics for L2 cache and off-chip DRAM in Table~\ref{tb:hw_config} since their configuration is identical between \name and our baselines.
% We will focus on reporting the statistics of one streaming multiprocessor and then scale the statistics to get the overall results for different types of GPU-like SIMT architectures.
%\footnote{We plan to open source our simulation codes to provide community a tool to quickly explore the design space of hyper-asymmetric GEMM.} 

\begin{table}[h]
\centering
\caption{Configuration of the PacQ and Baselines.}
\vspace{-3mm}
\begin{adjustbox}{max width =0.9\linewidth}
\begin{tabular}{l | l}
\toprule
INT11 MUL (baseline)& 10 INT16 adders\\
Parallel INT11 MUL& 12 INT16 adders, 4 INT6 adders\\
\midrule
FP16 MUL (baseline)& 1 INT11 MUL, 1 INT5 adder\\
& 1 normalization unit, 1 rounding unit\\
Parallel FP-INT-16 MUL & 1 parallel INT11 MUL, 1 INT5 adder\\
 & 1 normalization unit, 4 rounding units \\
\midrule
FP-16 DP-4 (baseline)& 4 FP16 MUL, 4 FP16 adders\\
Parallel FP-INT-16 DP-4 & 4 parallel FP-INT-16 MUL, 8 FP16 adders\\
\midrule
Tensor Core & 4 Parallel FP-INT-16 DP-4 \\
& baseline: 4 FP16 DP-4\\
& 2$\times$3072-bits buffer, 256KB register files\cite{v100whitepaper}\\
Streaming Multiprocessor & 8 tensor cores, 96 KB shared L1 cache\\
\bottomrule
\end{tabular}
\end{adjustbox}
\label{tb:hw_config}
\end{table}

% \subsection{}

\begin{figure}[h]
\centering
\includegraphics[width=\linewidth]{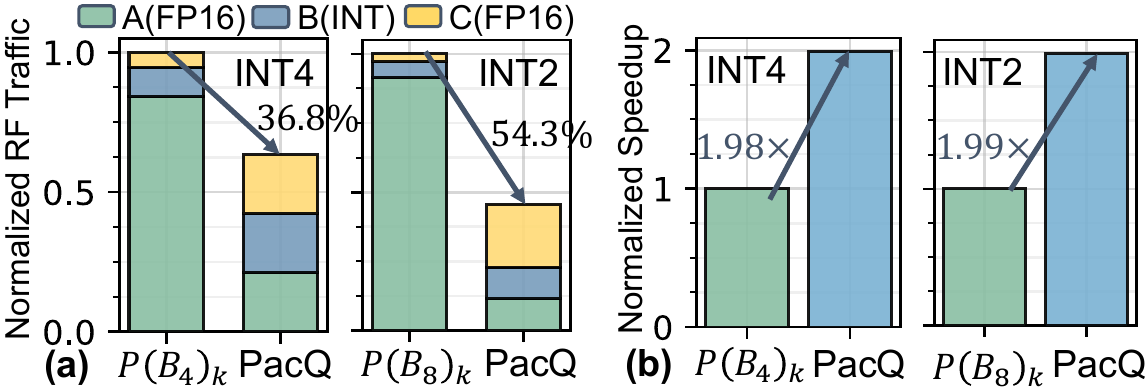}
\vspace{-5mm}
% \caption{(a) Normalized number of register file accesses comparison. (b) Normalized speedup comparison. The workload is \textit{m16n16k16}. $P(B_{4(8)})_k$ denotes packing 4 INT4 weights(8 INT2 weights) in one INT16 data along the $k$ dimension.}
\caption{(a) Comparison of the normalized number of register file accesses. (b) Comparison of normalized speedup between \name and hyper-asymmetric GEMM with INT weights packed along $k$. The workload is \textit{m16n16k16}. $P(B_{4(8)})_k$ represents packing 4 INT4 weights (or 8 INT2 weights) into one INT16 data along the $k$ dimension.}
  \label{fig:result:rf_access}
  \vspace{-3mm}
\end{figure}

\noindent \textbf{Proposed Packing and Data Flow.} \indent In Figure.~\ref{fig:result:rf_access}(a), we present the number of register file accesses for \name compared to the hyper-asymmetric GEMM with INT4 (INT2) weights packed along $k$. The results demonstrate that with improved data reuse, our approach achieves up to a $54.3\%$ reduction in register file accesses. Furthermore, in Figure.~\ref{fig:result:rf_access}(b), we show the normalized speedup of \name compared to the hyper-asymmetric GEMM packing along $k$. Thanks to our parallel processing strategy, we achieve an average speedup of $1.99\times$.

% \vspace{-3mm}
\begin{table}[h]
    \centering
% \caption{Perplexity of RTN-based PTQ Llama2 models~\cite{touvron2023llama} on WikiText-2 and C4. W4A16-g128 means 4-bit weight-only quantization with a 128-gourp size along $k$ dimension. g[32,4] means 128-group size with 32 groups along $k$ and 4 along $n$.}
\caption{Perplexity of RTN-based PTQ Llama2 models on WikiText-2 and C4. W4A16-g128 refers to 4-bit weight-only quantization with a 128-group size along $k$-dim. g[32,4] indicates a 128-group size distributed as 32 groups along $k$ and 4 groups along $n$.}
    \vspace{-2mm}
    \begin{adjustbox}{max width=\linewidth}
    \begin{tabular}{cccccc}
        \toprule
        Model&W16A16&W4A16 & W4A16 & W4A16 & W4A16\\
        Llama2-7B &fp16 baseline& g128 & g[32,4] & g256 & g[64,4]\\
        \midrule
        \midrule
         wikitext-2&5.47& 5.73 & 5.72&5.75&5.77\\
       C4&7.26& 7.58 & 7.59& 7.64&7.66\\
        \bottomrule
    \end{tabular}
    \end{adjustbox}
    \label{tab:quant:ppl}
\end{table}

% \noindent We want to confirm here that \name \textit{\textbf{does not require any quantization algorithm modifications}} to gain the hardware efficiency, since there is no approximation in our design. However, we do observe that small change to the existing PTQ algorithm can help improve the efficiency of \name. More specifically, if we span the quantization group on both dimensions ($[n,k]$) instead of just on input feature dimension ($k$), we can reduce the number of fetches of quantization scale $s$ for \name in the general core, as illustrated in Figure.~\ref{fig:arch}-\circled{3}. We modify the standard round-to-nearest (RTN) based PTQ algorithm to have quantization group on both $n$ and $k$ dimension in the existing framework~\cite{llmc} and show the perplexity across two datasets for Llama2-7B in Table.~\ref{tab:quant:ppl}. Results show that with the quantization group-dimension modifications, we achieve the iso-perplexity with standard RTN 4-bit weight-only quantized models.
 We want to confirm here that \name \textit{\textbf{does not require any quantization algorithm modifications}} to achieve hardware efficiency, as there is no approximation in our design. However, we observe that a small adjustment to the existing PTQ algorithm can further enhance the efficiency of \name. Specifically, by spanning the quantization group across both dimensions ($[n, k]$) instead of solely on the input feature dimension ($k$), we can reduce the number of fetches of the quantization scale $s$ for \name in the general core, as illustrated in Figure~\ref{fig:arch}-\circled{3}. We adapt the standard round-to-nearest (RTN) based PTQ algorithm to define quantization groups across both $n$ and $k$ dimensions within the existing framework~\cite{llmc}. The perplexity results for Llama2-7B~\cite{touvron2023llama} are shown in Table~\ref{tab:quant:ppl}. The results demonstrate that with these quantization group-dimension modifications, we achieve iso-perplexity with standard RTN 4-bit weight-only quantized models.

% \subsection{}

%%%% Figure 1 %%%%
\begin{figure}[h]
\centering
\includegraphics[width=\linewidth]{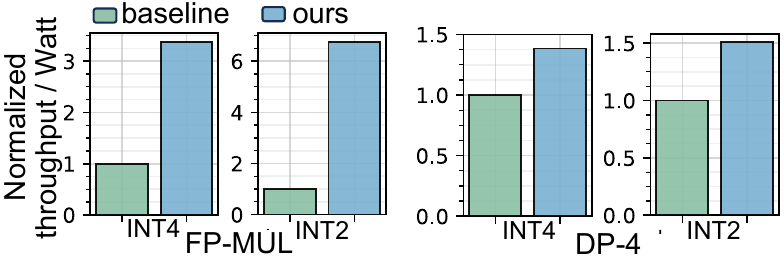}
\vspace{-5mm}
  \caption{Normalized performance  comparison between the baseline FP16 designs with our parallel FP-INT-16 designs of running INT4 and INT2 weights. For DP-4, we consider the workload of \textit{m2n4k4}.}
  \label{fig:result:throughput_watt}
\end{figure}
%%%% Figure 1 %%%%

% \noindent \textbf{Parallel FP-INT Multiplier.}\indent We first use the metric of throughput per watt to indicate the overall performance of our proposed parallel FP-INT multiplier. In Figure.~\ref{fig:result:throughput_watt}, we compare the normalized performance between our design and the original FP16 multiplier and FP16 DP-4 unit. 
% The throughput of the origin FP16 multiplier is 1 and 4 (8) for our parallel design with INT4 (INT2). The origin FP16 DP4 unit requires 11 cycles to generate 8 FP16 outputs, while our parallel design requires 19 (35) cycles to generate 32 (64) FP16 outputs for INT4 (INT2) weight. Compared to the origin FP16 multiplier, our design achieves 3.38$\times$ better performance for INT4 and 6.75$\times$ for INT2.
\noindent \textbf{Parallel FP-INT Multiplier.}\indent We first evaluate the throughput / Watt to measure the overall performance of our proposed parallel FP-INT multiplier. Figure~\ref{fig:result:throughput_watt} compares the normalized performance of our design against the original FP16 multiplier and the FP16 DP-4 unit.
The throughput of the original FP16 multiplier is 1, whereas it is 4 (8) for our parallel design with INT4 (INT2) weights. The original FP16 DP-4 unit requires 11 cycles to generate 8 FP16 outputs, while our parallel design takes 19 (35) cycles to generate 32 (64) FP16 outputs for INT4 (INT2) weights. Compared to the original FP16 multiplier, our design achieves 3.38$\times$ better performance for INT4 weights and 6.75$\times$ better performance for INT2.

%%%% Figure 1 %%%%
\begin{figure}[h]
\centering
\includegraphics[width=0.9\linewidth]{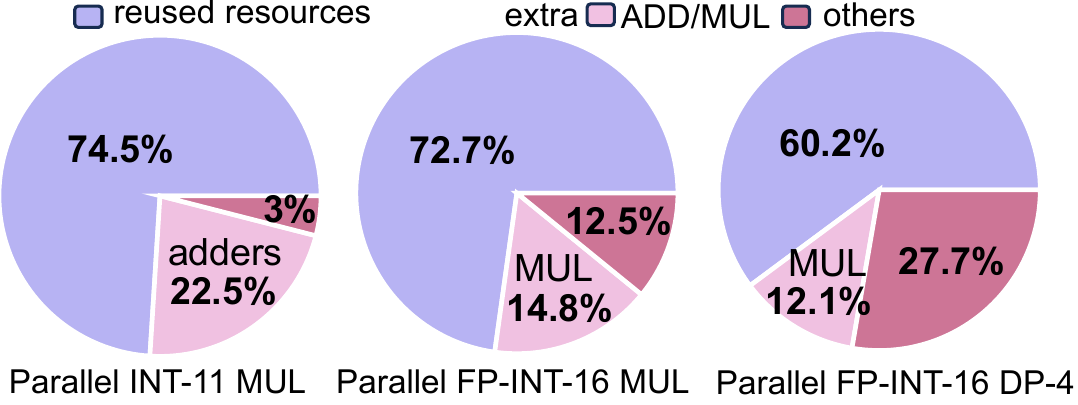}
\vspace{-3mm}
  \caption{Power breakdown of our proposed units. MUL stands for multiplier.}
  \label{fig:result:power:breakup}
\end{figure}
%%%% Figure 1 %%%%
% We further show the power breakdown of our proposed units in Figure.~\ref{fig:result:power:breakup}. Our design goal is to reuse as much hardware resources from the standard design as possible. As shown in the purple colored portion, we have managed to reuse almost 75\% of the original INT-11 multiplier resources, which is the most basic building block in our parallel FP-INT multiplier. We end up having around 60\% of hardware resource reuse in the DP-4 unit. The trend of a constant decrease in the resource reuse ratio is due to the duplication of other units (e.g., we need to double the FP16 adder trees in DP-4). Overall, we get on average 69\% of hardware resource reuse ratio, which largely saves our design from wasting on duplication units.

\noindent We further show the power breakdown of our proposed units in Figure~\ref{fig:result:power:breakup}. Our design objective is to maximize the reuse of hardware resources from the standard design. As shown by the \textcolor[HTML]{6C74C4}{purple-colored} portions in the figure, we successfully reuse nearly 75\% of the original INT-11 multiplier resources, which serve as the fundamental building blocks in our parallel FP-INT multiplier. For the DP-4 unit, we achieve approximately 60\% hardware resource reuse. The observed decrease in the resource reuse ratio stems from the duplication of additional units; for instance, we needed to double the FP16 adder trees in the DP-4 design. Despite this, our design maintains an average hardware resource reuse ratio of 69\%, which significantly minimizes resource duplication and enhances overall efficiency.

% \subsection{}
\begin{figure}[h]
\centering
\includegraphics[width=0.8\linewidth]{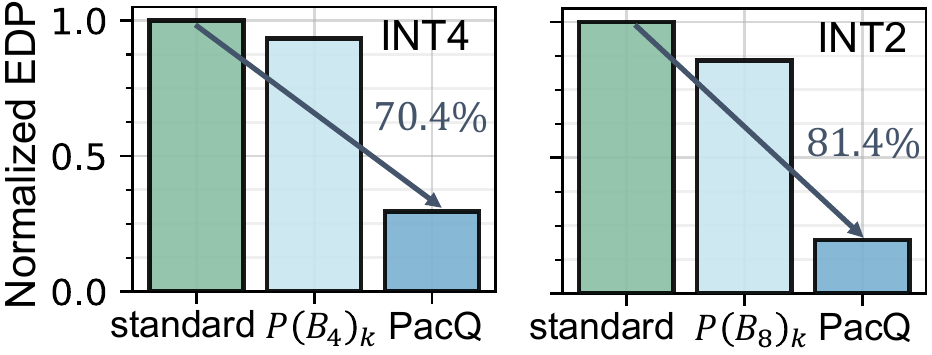}
\vspace{-3mm}
% \caption{Normalized EDP comparison of our \name with baselines. 'standard' is the standard dequantization-based W16A16 GEMM. $P(B_{4(8)})_k$ denotes the hyper-asymmetric GEMM with INT4(INT2) weights packed along $k$.}
\caption{Normalized EDP of \name vs. baselines. 'Standard' refers to the standard dequantization-based W16A16 GEMM. $P(B_{4(8)})_k$ denotes the hyper-asymmetric GEMM with INT4 (INT2) weights packed along $k$.}
  \label{fig:result:energy_speedup}
\end{figure}

% \noindent \textbf{Other Results.}\indent In Figure.~\ref{fig:result:energy_speedup}, we show the overall energy-delay-product (EDP) of \name compared to the Volta-ike SIMT baselines. We compare with two cases: the standard dequantization-based GEMM flow and the hyper-asymmetric GEMM flow with the packing along $k$ dimension. Result shows that \name can reduce up to $81.4\%$ EDP reduction on the workload of \textit{m}16\textit{n}16384\textit{k}4096, which is a basic GEMM workload in Llama2-7B with 16 batches.

\noindent \textbf{Other Results.}\indent In Figure.~\ref{fig:result:energy_speedup}, we present the overall energy-delay-product (EDP) comparison of \name against Volta-like SIMT baselines. Two baseline cases are evaluated: the standard dequantization-based GEMM flow and the hyper-asymmetric GEMM flow with weights packed along the $k$ dimension. Results demonstrate that \name achieves up to an $81.4\%$ reduction in EDP for the workload \textit{m}16\textit{n}4096\textit{k}4096, which represents a FFN layer in Llama2-7B with 16 batches.

% \subsection{and Analyses}

%%%% Figure 1 %%%%
\begin{figure}[h]
\centering
\includegraphics[width=0.75\linewidth]{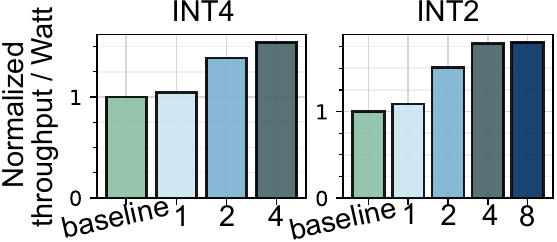}
  \vspace{-3mm}
  % \caption{Ablation study on the adder tree duplication. Number indicates the level of duplication of the adder trees (total 4 FP16 adders in baseline). For example, 4 would indicate 16 FP16 adders in our parallel FP-INT-16 DP-4.}
  \caption{Ablation study on the adder tree duplication. The number indicates the level of duplication of the adder trees (with a total of 4 FP16 adders in the baseline). For example, a duplication level of 4 would indicate 16 FP16 adders in our parallel FP-INT-16 DP-4 design.}
  \vspace{-3mm}
  \label{fig:result:ablation:duplication}
\end{figure}
%%%% Figure 1 %%%%
 % In Figure.~\ref{fig:result:ablation:duplication}, we examine how the number of duplication of FP16 adder trees in our parallel FP-INT DP-4 unit affect the overall performance (throughput / watt) on \textit{m16n16k16} workload. We observe that while increasing the duplication level will keep improving the overall performance, duplication of 2 gives the best gain. It improves 1.33$\times$ (1.38$\times$) from duplication of 1 for INT4 (INT2). Comparatively, duplication of 4 only improves 1.11$\times$ (1.18$\times$) from duplication of 2 for INT4 (INT2). 

\noindent In Figure.~\ref{fig:result:ablation:duplication}, we investigate the impact of duplicating FP16 adder trees in our parallel FP-INT DP-4 unit on overall performance (throughput / Watt) for the \textit{m16n16k16} workload. We observe that while increasing the level of duplication continues to improve performance, a duplication factor of 2 provides the most significant gain. It achieves a 1.33$\times$ (1.38$\times$) improvement over a duplication factor of 1 for INT4 (INT2). In comparison, a duplication factor of 4 results in only a 1.11$\times$ (1.18$\times$) improvement over duplication factor 2 for INT4 (INT2).
 
\begin{figure}[h]
\centering
\includegraphics[width=0.85\linewidth]{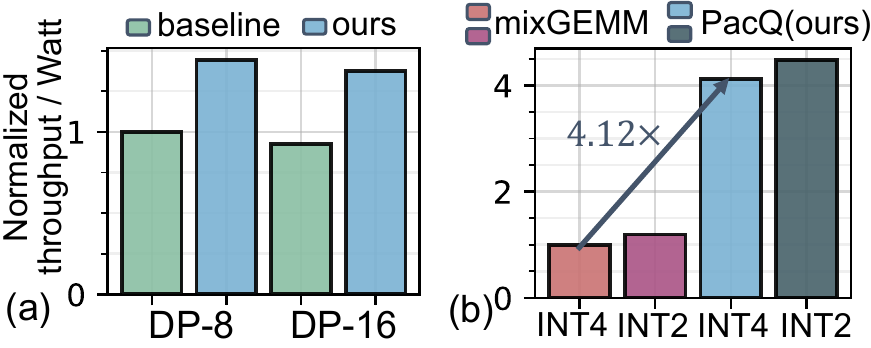}
\vspace{-3mm}
% \caption{(a) Comparison of different size of DP unit. (b) Comparison of \name with Mix-GEMM~\cite{reggiani2023mix}. Both (a) and (b) use metric of throughput/watt. And the workloads are both \textit{m16n16k16}.}
\caption{(a) Comparison of different sizes of the DP unit. (b) Comparison of \name with Mix-GEMM~\cite{reggiani2023mix}. Both (a) and (b) use the metric of throughput per watt. The workloads for both comparisons are \textit{m16n16k16}.}
  \label{fig:result:priorwork_comparison}
\end{figure}

 % In Figure.~\ref{fig:result:priorwork_comparison}(a), we further studies the effect of the size of DP-unit (DP-8 and DP-16). The result shows that \name's performance improvements is orthogonal to the size of DP units. In Figure.~\ref{fig:result:priorwork_comparison}(b), we compare the performance of \name with the prior work of Mix-GEMM~\cite{reggiani2023mix} which also targets to improve the throughput of the mixed-precision GEMM workloads through binary-segmentation. With FP16 activation, \name achieves $4.12\times$ better performance for INT4 weights and $3.75\times$ for INT2. The reason is that binary segmentation technique performs poorly for hyper-asymmetric GEMM.

 \noindent In Figure.~\ref{fig:result:priorwork_comparison}(a), we further study the effect of the size of the DP unit (DP-8 and DP-16). The results show that \name's performance improvements are orthogonal to the size of the DP units. In Figure.~\ref{fig:result:priorwork_comparison}(b), we compare the performance of \name with prior work, Mix-GEMM~\cite{reggiani2023mix}, which also aims to improve the throughput of mixed-precision GEMM workloads through binary segmentation. With FP16 activation, \name achieves a $4.12\times$ performance improvement for INT4 weights and a $3.75\times$ improvement for INT2. This performance advantage is attributed to the fact that the binary segmentation technique performs poorly for hyper-asymmetric GEMM.

\section{Conclusion}
In this work, we present \name, a novel SIMT microarchitecture designed to efficiently handle hyper-asymmetric GEMMs with mixed-precision weights (INT4 and INT2) and activations (FP16). By introducing a new packing and dataflow strategy, coupled with our parallel FP-INT multiplier, \name achieves significant improvements in hardware efficiency and throughput. Through extensive experimental evaluation, we demonstrate that \name delivers up to $54.3\%$ reduction in register file accesses, a $1.99\times$ speedup in throughput, and up to $81.4\%$ reduction in energy-delay product (EDP) compared to conventional SIMT baselines. We believe our work can provide the community with insights into the design space exploration for hyper-asymmetric GEMM acceleration.

\section{Acknowledgement}
This work was supported in part by CoCoSys, a JUMP2.0 center sponsored by DARPA and SRC, the National Science Foundation (CAREER Award, Grant \#2312366, Grant \#2318152), the DARPA Young Faculty Award and the DoE MMICC center SEA-CROGS (Award \#DE-SC0023198).
%%%%%%%%% -- BIB STYLE AND FILE -- %%%%%%%%
\bibliographystyle{IEEEtranS}
\bibliography{refs}

% Generated by IEEEtranS.bst, version: 1.14 (2015/08/26)
\begin{thebibliography}{10}
\providecommand{\url}[1]{#1}
\csname url@samestyle\endcsname
\providecommand{\newblock}{\relax}
\providecommand{\bibinfo}[2]{#2}
\providecommand{\BIBentrySTDinterwordspacing}{\spaceskip=0pt\relax}
\providecommand{\BIBentryALTinterwordstretchfactor}{4}
\providecommand{\BIBentryALTinterwordspacing}{\spaceskip=\fontdimen2\font plus
\BIBentryALTinterwordstretchfactor\fontdimen3\font minus \fontdimen4\font\relax}
\providecommand{\BIBforeignlanguage}[2]{{%
\expandafter\ifx\csname l@#1\endcsname\relax
\typeout{** WARNING: IEEEtranS.bst: No hyphenation pattern has been}%
\typeout{** loaded for the language `#1'. Using the pattern for}%
\typeout{** the default language instead.}%
\else
\language=\csname l@#1\endcsname
\fi
#2}}
\providecommand{\BIBdecl}{\relax}
\BIBdecl

\bibitem{autogptq}
A.~contributors, ``Autogptq,'' \url{https://github.com/AutoGPTQ}, 2024.

\bibitem{frantar2022gptq}
E.~Frantar, S.~Ashkboos, T.~Hoefler, and D.~Alistarh, ``Gptq: Accurate post-training quantization for generative pre-trained transformers,'' \emph{arXiv preprint arXiv:2210.17323}, 2022.

\bibitem{frantar2022optq}
------, ``Optq: Accurate quantization for generative pre-trained transformers,'' in \emph{The Eleventh International Conference on Learning Representations}, 2022.

\bibitem{gope2020high}
D.~Gope, J.~Beu, and M.~Mattina, ``High throughput matrix-matrix multiplication between asymmetric bit-width operands,'' \emph{arXiv preprint arXiv:2008.00638}, 2020.

\bibitem{jacob2018quantization}
B.~Jacob, S.~Kligys, B.~Chen, M.~Zhu, M.~Tang, A.~Howard, H.~Adam, and D.~Kalenichenko, ``Quantization and training of neural networks for efficient integer-arithmetic-only inference,'' in \emph{Proceedings of the IEEE conference on computer vision and pattern recognition}, 2018, pp. 2704--2713.

\bibitem{jang2024figna}
J.~Jang \emph{et~al.}, ``Figna: Integer unit-based accelerator design for fp-int gemm preserving numerical accuracy,'' in \emph{2024 IEEE International Symposium on High-Performance Computer Architecture (HPCA)}.\hskip 1em plus 0.5em minus 0.4em\relax IEEE, 2024, pp. 760--773.

\bibitem{kim2022says}
Y.~J. Kim, R.~Henry, R.~Fahim, and H.~H. Awadalla, ``Who says elephants can't run: Bringing large scale moe models into cloud scale production,'' \emph{arXiv preprint arXiv:2211.10017}, 2022.

\bibitem{li2021brecq}
Y.~Li, R.~Gong, X.~Tan, Y.~Yang, P.~Hu, Q.~Zhang, F.~Yu, W.~Wang, and S.~Gu, ``Brecq: Pushing the limit of post-training quantization by block reconstruction,'' \emph{arXiv preprint arXiv:2102.05426}, 2021.

\bibitem{li2024tesseraq}
Y.~Li and P.~Panda, ``Tesseraq: Ultra low-bit llm post-training quantization with block reconstruction,'' \emph{arXiv preprint arXiv:2410.19103}, 2024.

\bibitem{lin2024awq}
J.~Lin \emph{et~al.}, ``Awq: Activation-aware weight quantization for on-device llm compression and acceleration,'' \emph{Proceedings of Machine Learning and Systems}, vol.~6, pp. 87--100, 2024.

\bibitem{llmc}
llmc contributors, ``llmc: Towards accurate and efficient llm compression,'' \url{https://github.com/ModelTC/llmc}, 2024.

\bibitem{mo2024lut}
Z.~Mo \emph{et~al.}, ``Lut tensor core: Lookup table enables efficient low-bit llm inference acceleration,'' \emph{arXiv preprint arXiv:2408.06003}, 2024.

\bibitem{cacti}
N.~Muralimanohar, R.~Balasubramonian, and N.~P. Jouppi, ``Cacti 6.0: A tool to model large caches,'' \emph{HP laboratories}, 2009.

\bibitem{v100whitepaper}
{NVIDIA Corporation}, ``{NVIDIA TESLA V100 GPU ARCHITECTURE},'' \url{http://images.nvidia.com/content/volta-architecture/pdf/volta-architecture-whitepaper.pdf}, June 2017.

\bibitem{park2022lut}
G.~Park, B.~Park, M.~Kim, S.~Lee, J.~Kim, B.~Kwon, S.~J. Kwon, B.~Kim, Y.~Lee, and D.~Lee, ``Lut-gemm: Quantized matrix multiplication based on luts for efficient inference in large-scale generative language models,'' \emph{arXiv preprint arXiv:2206.09557}, 2022.

\bibitem{demystifyvolta}
M.~A. Raihan \emph{et~al.}, ``Modeling deep learning accelerator enabled gpus,'' in \emph{2019 IEEE International Symposium on Performance Analysis of Systems and Software (ISPASS)}.\hskip 1em plus 0.5em minus 0.4em\relax IEEE, 2019, pp. 79--92.

\bibitem{reggiani2023mix}
E.~Reggiani \emph{et~al.}, ``Mix-gemm: An efficient hw-sw architecture for mixed-precision quantized deep neural networks inference on edge devices,'' in \emph{2023 IEEE International Symposium on High-Performance Computer Architecture (HPCA)}.\hskip 1em plus 0.5em minus 0.4em\relax IEEE, 2023, pp. 1085--1098.

\bibitem{tan2011fast}
G.~Tan \emph{et~al.}, ``Fast implementation of dgemm on fermi gpu,'' in \emph{Proceedings of 2011 International Conference for High Performance Computing, Networking, Storage and Analysis}, 2011, pp. 1--11.

\bibitem{touvron2023llama}
H.~Touvron, T.~Lavril, G.~Izacard, X.~Martinet, M.-A. Lachaux, T.~Lacroix, B.~Rozi{\`e}re, N.~Goyal, E.~Hambro, F.~Azhar \emph{et~al.}, ``Llama: Open and efficient foundation language models,'' \emph{arXiv preprint arXiv:2302.13971}, 2023.

\bibitem{wang2023bitnet}
H.~Wang, S.~Ma, L.~Dong, S.~Huang, H.~Wang, L.~Ma, F.~Yang, R.~Wang, Y.~Wu, and F.~Wei, ``Bitnet: Scaling 1-bit transformers for large language models,'' \emph{arXiv preprint arXiv:2310.11453}, 2023.

\bibitem{xiao2023smoothquant}
G.~Xiao, J.~Lin, M.~Seznec, H.~Wu, J.~Demouth, and S.~Han, ``Smoothquant: Accurate and efficient post-training quantization for large language models,'' in \emph{International Conference on Machine Learning}.\hskip 1em plus 0.5em minus 0.4em\relax PMLR, 2023, pp. 38\,087--38\,099.

\bibitem{yu2022orca}
G.-I. Yu, J.~S. Jeong, G.-W. Kim, S.~Kim, and B.-G. Chun, ``Orca: A distributed serving system for $\{$Transformer-Based$\}$ generative models,'' in \emph{16th USENIX Symposium on Operating Systems Design and Implementation (OSDI 22)}, 2022, pp. 521--538.

\bibitem{zhang2022opt}
S.~Zhang, S.~Roller, N.~Goyal, M.~Artetxe, M.~Chen, S.~Chen, C.~Dewan, M.~Diab, X.~Li, X.~V. Lin \emph{et~al.}, ``Opt: Open pre-trained transformer language models,'' \emph{arXiv preprint arXiv:2205.01068}, 2022.

\bibitem{zhao2024atom}
Y.~Zhao, C.-Y. Lin, K.~Zhu, Z.~Ye, L.~Chen, S.~Zheng, L.~Ceze, A.~Krishnamurthy, T.~Chen, and B.~Kasikci, ``Atom: Low-bit quantization for efficient and accurate llm serving,'' \emph{Proceedings of Machine Learning and Systems}, vol.~6, pp. 196--209, 2024.

\end{thebibliography}
%%%%%%%%%%%%%%%%%%%%%%%%%%%%%%%%%%%%
\end{document}